\documentclass{emulateapj}
\usepackage{epsfig}
\makeatletter

\begin{document}

    \newcommand{\fermi}{{\it Fermi}}
	\newcommand{\chandra}{{\it Chandra}}
	\newcommand{\hst}{{\it HST}}
	\newcommand{\asca}{{\it ASCA}}
	\newcommand{\rosat}{{\it ROSAT}}
	\newcommand{\sax}{{\it BeppoSAX}}
	\newcommand{\xmm}{{\it XMM-Newton}}
	\newcommand{\swift}{{\it Swift}}
	\newcommand{\lum}{\thinspace\hbox{$\hbox{erg}\thinspace\hbox{s}^{-1}$}}
	\newcommand{\flux}{\thinspace\hbox{$\hbox{erg}\thinspace\hbox{cm}^{-2}\thinspace\hbox{s}^{-1}$}}
\slugcomment{Accepted for publication in ApJL, February 11, 2010}

	\title{\fermi\ Discovery of Gamma-ray Emission from the Globular Cluster Terzan 5}
	\author{A.~K.~H.~Kong\altaffilmark{1,4}, C.~Y. Hui\altaffilmark{2}, and K.~S. Cheng\altaffilmark{3}}
	\altaffiltext{1}{Institute of Astronomy and Department of Physics, National Tsing Hua University, Hsinchu 30013, Taiwan; akong@phys.nthu.edu.tw}
	\altaffiltext{2}{Department of Astronomy and Space Science, Chungnam National University, Daejeon, South Korea}
	\altaffiltext{3}{Department of Physics, The University of Hong Kong, Hong Kong, China}
	\altaffiltext{4}{Kenda Foundation Golden Jade Fellow}

\begin{abstract}
We report the discovery of gamma-ray emission from the Galactic globular cluster Terzan 5 using data taken with the {\it Fermi Gamma-ray Space Telescope}, from 2008 August 8 to 2010 January 1. Terzan 5 is clearly detected in the 0.5--20 GeV band by \fermi\ at $\sim 27\sigma$ level. This makes Terzan 5 as the second gamma-ray emitting globular cluster seen by \fermi\ after 47 Tuc. The energy spectrum of Terzan 5 is best represented by an exponential cutoff power-law model, with a photon index of $\sim 1.9$ and a cutoff energy at $\sim 3.8$ GeV. By comparing to 47 Tuc, we suggest that the observed gamma-ray emission is associated with millisecond pulsars, and is either from the magnetospheres or inverse Compton scattering between the relativistic electrons/positrons in the pulsar winds and the background soft photons from the Galactic plane. Furthermore, it is suggestive that the distance to Terzan 5 is less than 10 kpc and $> 10$ GeV photons can be seen in the future.

\end{abstract}


\section{Introduction}
Owing to the high stellar densities in globular clusters (GCs), GCs are effectively
pulsar factories with the production through frequent dynamical interactions. One 
should notice that more than $\sim80\%$ of detected millisecond pulsars (MSPs) are 
located in GCs. Thanks to the extensive radio surveys, 140 MSPs have been identified 
in 26 GCs so far.\footnote{http://www2.naic.edu/$\sim$pfreire/GCpsr.html} 

The radio and X-ray properties of MSPs in GCs are found to be rather different 
from those located in the Galactic field (Bogdanov et al. 2006; Hui, Cheng \& 
Taam 2009). The difference can be possibly related to the complicated magnetic 
field structure of the MSPs in a cluster, which is a consequence of frequent 
stellar interaction (Cheng \& Taam 2003). Gamma-ray observations can lead us to 
a deeper insight of whether the high energy emission processes and hence the 
magnetospheric structure are fundamentally different between these two populations. 
However, this regime has not been fully explored until very recently. 

With the launch of the {\it Fermi Gamma-ray Space Telescope (Fermi)}, we have entered a new 
era of high energy astrophysics. As the sensitivity of the Large Area Telescope 
(LAT) onboard \fermi\ is much higher than that of its predecessor, EGRET, it has already 
led to many interesting discoveries. Shortly after its operation, the gamma-ray 
emission from 47~Tuc in MeV$-$GeV regime have been detected with high significance 
(Abdo et al. 2009a). Since the other types of close binaries in a cluster, such as 
catalysmic variables and low-mass X-ray binaries, have not yet been detected in 
gamma-ray (Abdo et al. 2009b), the gamma-rays detected from 47~Tuc are 
presumably contributed by its MSP population. There are 23 MSPs have been uncovered 
in 47~Tuc so far. Both X-ray and gamma-ray analysis have placed an upper bound of 
$\sim60$ on its possible MSP population. Venter, De Jager \& Clapson (2009) suggest 
that curvature radiation results directly from the MSP magnetospheres should be 
the dominant contributor of the observed gamma-rays in MeV--GeV regime. Apart from  
this component, the Inverse Compton scattering between the surrounding soft photon 
field and the energetic electrons/positrons from the pulsars can also have an additional 
contribution to the gamma-ray emission (Cheng et al. in preparation).  

While 47~Tuc is located relatively close to us, $\sim4.0-4.5$~kpc (Harris 1996; 
McLaughlin et al. 2006), Terzan~5 holds the largest MSP population among all 
the MSP-hosting GCs. Currently, there are 33 pulsars have been found in Terzan~5. 
Based on the radio luminosity functions of the MSPs in GCs reported by Hui, Cheng \& Taam (2010), the hidden pulsar population in Terzan~5 is found to be much larger 
than that of 47~Tuc. By scaling the upper limit of 60 MSPs in 47~Tuc to a corresponding 
limit in Terzan~5 with the aids of the luminosity functions, we found that the number 
of pulsars in Terzan~5 can be as large as $\sim200$ which is $\sim3$ times larger 
than that in 47~Tuc. In view of this large predicted population, the collective 
gamma-ray intensity from the MSPs in Terzan~5 are expected to be well above 
the detection threshold with more than one-year long LAT data. In this Letter, 
we report the search and the analysis of the gamma-ray emission from Terzan~5 using \fermi\ LAT. 

\begin{figure*}
	\centering
	\psfig{file=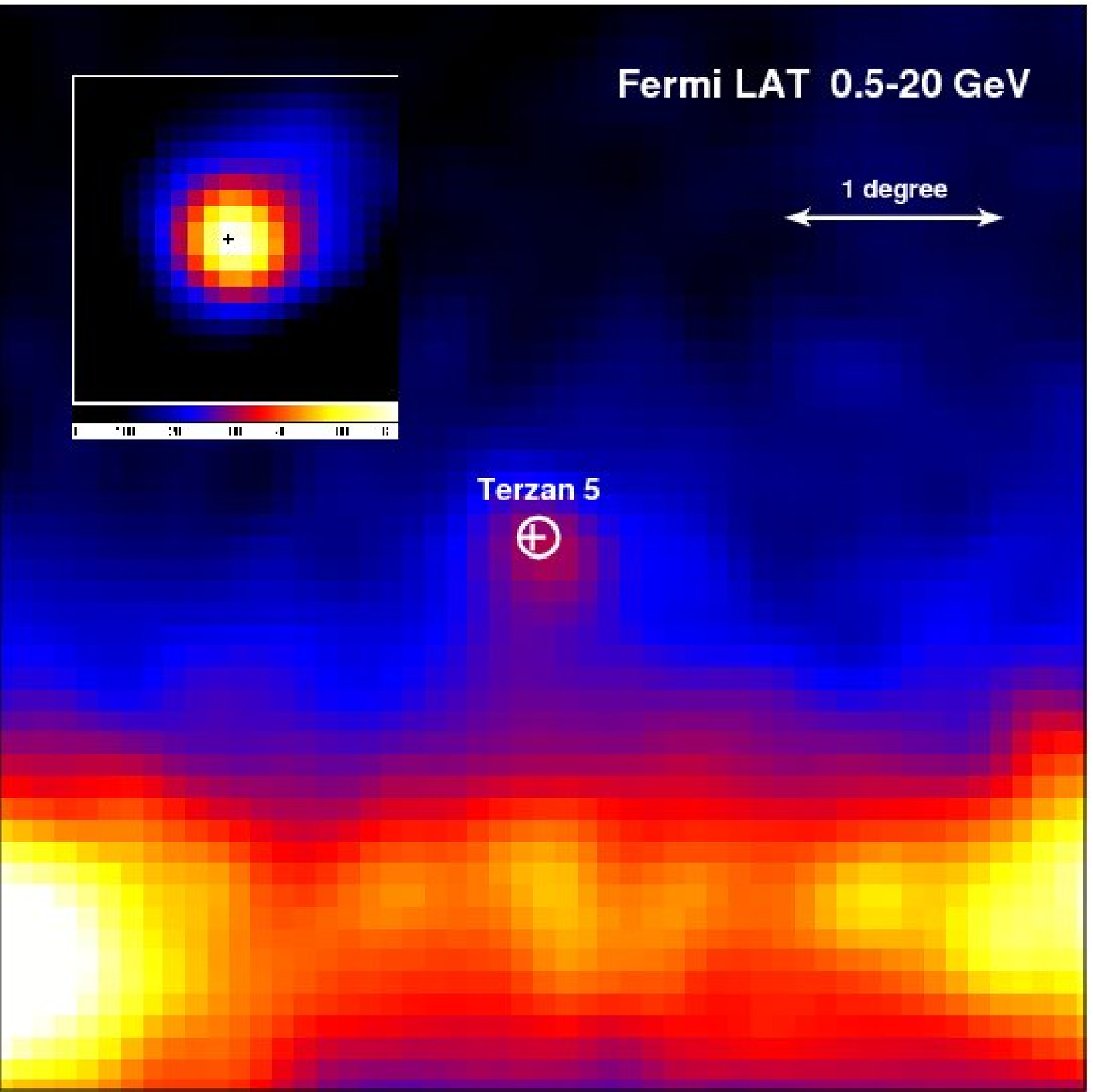,width=3.4in}
	\psfig{file=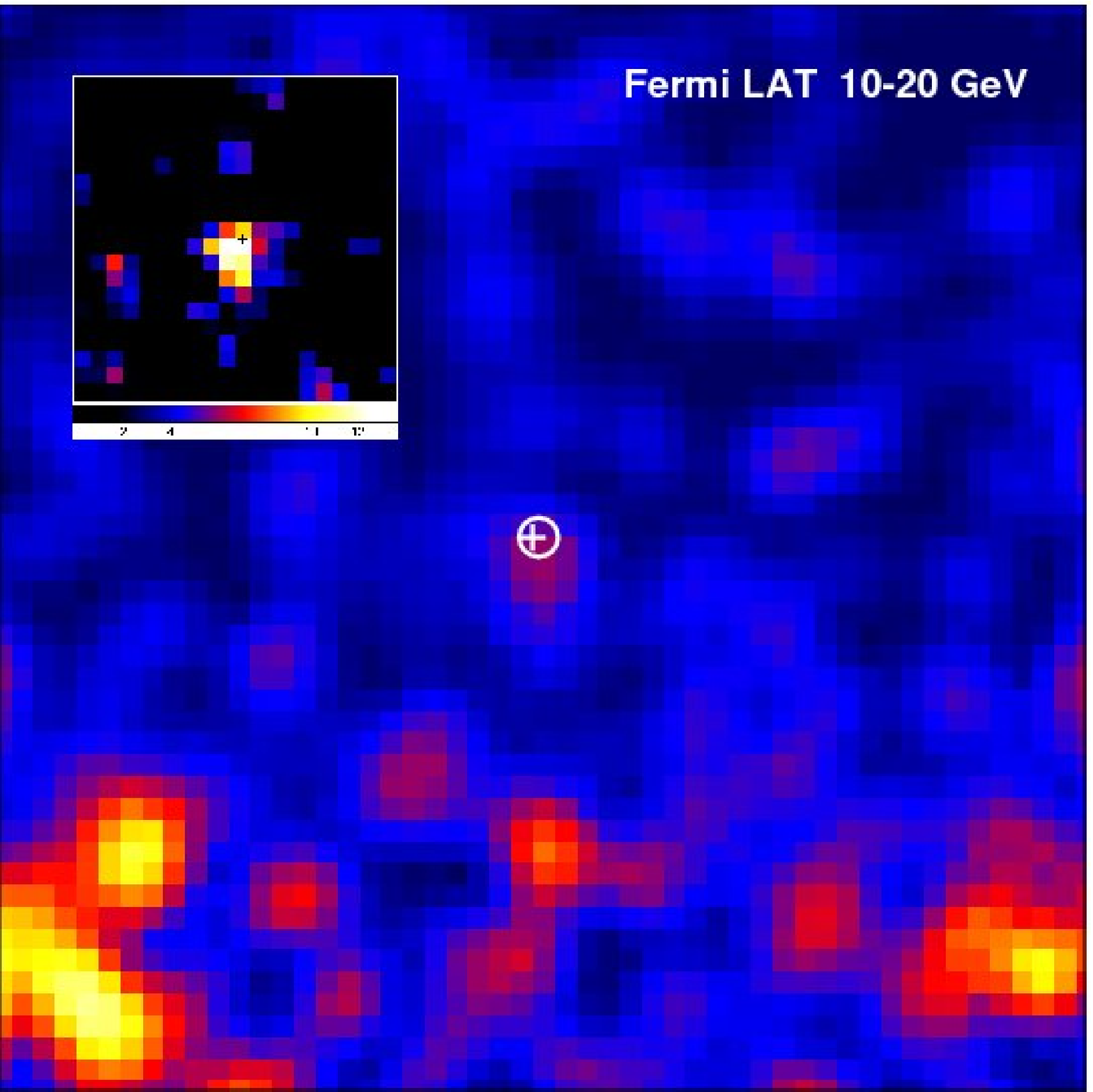, width=3.4in}
	\caption{\fermi\ LAT 0.5--20 GeV (left) and 10--20 GeV (right) images of a $5^\circ \times 5^\circ$ region centered on Terzan 5. The white circle shows the 95\% confidence error circle of the gamma-ray source determined by using the 0.5--20 GeV image and {\it gtfindsrc}. The white crosses are the optical center of Terzan 5. In addition to Terzan 5, strong Galactic plane emission is also seen. Both images are smoothed by a $0.3^\circ$ Gaussian function. The insert figures are the associated TS maps of the $2^\circ \times 2^\circ$ region. All bright \fermi\ sources as well as the diffuse Galactic and extragalactic background emission are included in the background model. The black crosses denote the optical center of Terzan 5. The color scale indicates the TS value corresponding to the detection significance $\sigma\approx\sqrt{TS}$. }
\end{figure*}

\section{Observations and Data Analysis}
The \fermi/LAT is a pair-production telescope that detects gamma-rays with energies between $\sim 20$ MeV and $> 300$ GeV (Atwood et al. 2009). It operates in an all-sky scanning mode which scans the whole sky in every 3 hours. In this study, we used the LAT data taken between 2008 August 4 and 2010 January 1. To reduce and analyze the data, we used the \fermi\ Science Tools v9r15p2 package and followed the data analysis threads provided by the \fermi\ Science Support 
Center\footnote{http://fermi.gsfc.nasa.gov/ssc/data/analysis/scitools/}. We selected data with ``Diffuse'' events, which have the highest probability of being a photon.
In addition, we filtered out events with earth zenith angles greater than $105^\circ$ to reduce the contamination from Earth albedo gamma-rays. The instrument response functions (IRFs), ``P6\_V3\_DIFFUSE'' are used. We limited our analysis using data between 0.5 and 20 GeV with which the point-spread-function is better than $\sim 1$ degree such that the contamination by the Galactic plane emission is minimized. Furthermore, the sensitivity is more uniform\footnote{http://www-glast.slac.stanford.edu/software/IS/glast\_lat\_performance.htm}.

In Figure 1, the photon map in the 0.5--20 GeV band in the vicinity of Terzan 5 is shown. Although strong Galactic plane emission is seen near Terzan 5, a gamma-ray source at the position of Terzan 5 is clearly detected as an isolated source at a level of $\sim 27\sigma$ (see below). We used {\it gtfindsrc} to determine the position of the gamma-ray position.
In order to minimize the contamination from nearby sources and diffuse emission, we included Galactic diffuse model (gll\_iem\_v02.fit) and isotropic background (isotropic\_iem\_v02.txt), as well as all point sources in the bright source catalog\footnote{http://fermi.gsfc.nasa.gov/ssc/data/access/lat/1yr\_catalog/} (Abdo et al. 2009b; version of 2010 January) within a region of interest of $15^\circ$ centered on Terzan 5.
The best-fit \fermi\ position of Terzan 5 is R.A.=17h48m00s, decl.=-24d48m15s (J2000) with a 95\% error of $0.09^\circ$ (including a 40\% systematic error based on the bright source catalog). The optical center of Terzan 5 is at R.A.=17h48m05s, decl.=-24d46m48s (J2000), which is $0.03^\circ$ from the \fermi\ position. Therefore the \fermi\ source is consistent with the location of Terzan 5. In addition, it is clear from the test-statistic (TS) map in Figure 1 that the source is highly significant in the 0.5--20 GeV band and the optical center of Terzan 5 falls on the peak of the TS map. Also shown in Figure 1 is the 10--20 GeV image and its TS map; the gamma-ray source has less than $4\sigma$ detection significance.
We also divided the image into four parts with roughly equal exposure time. Terzan 5 is always seen with no apparent flaring, indicating that the gamma-rays are not from any kind of flares.

We performed spectral analysis by using maximum likelihood technique implemented by {\it gtlike}. Like spatial analysis, we also included diffuse emission and bright point sources to obtain a simultaneous fit to the unbinned data. All bright point sources were modeled with simple power-laws. We first fit the spectrum of Terzan 5 with a simple power-law; the best-fitting photon index $\Gamma$ is $2.5\pm0.1$ (errors are statistical only). However, as shown by the gamma-ray emission from 47 Tuc as well as other MSPs, the energy spectrum is likely to be in the form of an exponential cutoff power-law (Abdo et al. 2009a,2009c). We show the \fermi\ spectrum of Terzan 5 in Figure 2 and it is clear that the energy distribution appears to turn over at energies above $\sim 5$ GeV. We then re-fit the spectrum with an exponential cutoff power-law model; the photon index is $\Gamma=1.9\pm0.2$ and the cutoff energy is $E_c=3.8\pm1.2$ GeV. The additional cutoff component is statistically significant at 99.99\% level. The resulting TS value, which indicates the significance of the source detection is 725 corresponding to $27\sigma$ significance.  We show the exponential cutoff power-law model together with the data in Figure 2. The 0.5--20 GeV photon flux of Terzan 5 is $(3.4\pm1.1)\times 10^{-8}$ photons cm$^{-2}$ s$^{-1}$, corresponding to $(6.8\pm2.0)\times 10^{-11}$ ergs cm$^{-2}$ s$^{-1}$. To compare with the reported 0.1--10 GeV flux of 47 Tuc (Abdo et al. 2009a), the 0.1--10 GeV flux of Terzan 5 is about $2\times 10^{-7}$ photons cm$^{-2}$ s$^{-1}$, corresponding to $1.2\times10^{-10}$ ergs cm$^{-2}$ s$^{-1}$. The photon flux of Terzan 5 is slightly below the 95\% upper limit of $2.6\times 10^{-7}$ photons cm$^{-2}$ s$^{-1}$ provided by EGRET (Michelson et al. 1994). 

\begin{figure}
	\centering
	\psfig{file=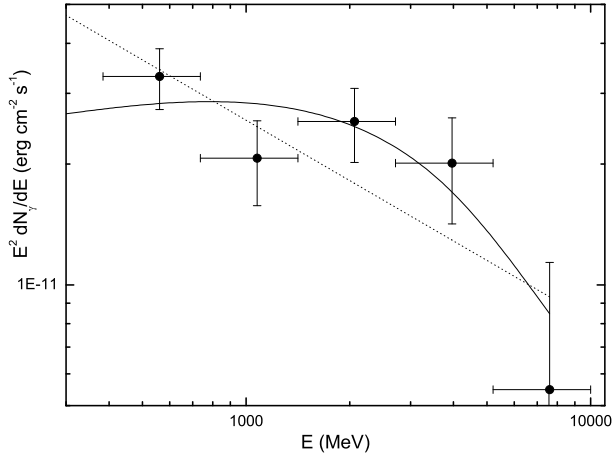,width=3.6in}
	\caption{\fermi\ LAT spectrum of Terzan 5. The solid line is the best-fit exponential cutoff power-law model obtained by {\it gtlike}. The dotted line is the best-fit simple power-law model. The error bars are statistical only.}
\end{figure}

\section{Discussion}
Using \fermi\ LAT data, we detect gamma-ray emission from Terzan 5 for the {\it first} time.
The gamma-rays emitted from GCs are generally assumed to associate with MSPs in clusters. The origin of these gamma-rays could be either pulsed curvature radiation arising near the polar cap and/or in outer magnetospheric gaps (e.g. Zhang \& Cheng 2003; Harding, Usov \& Muslimov 2005; Venter \& De Jager 2008), or inverse Compton scattering photons between the relativistic electrons/positrons in the pulsar winds and the background soft photons (e.g. Bednarek \& Sitarek 2007). 

Let us first assume that these gamma-rays have the magnetospheric origin.  It is useful to compare Terzan 5 with 47 Tuc, which is the first globular cluster detected with gamma-rays (Abdo et al. 2009a). The logarithm of central luminosity density and core radius for Terzan 5 and 47 Tuc are 5.06 $L_{\odot}pc^{-3}$ and 4.81 $L_{\odot}pc^{-3}$, and 0.54$pc$ and 0.52$pc$ respectively (Harris 1996; version of 2003 February). However the two body encounter rate and metallicity of Terzan 5 are higher than those of 47 Tuc. Since these two quantities favor the formation of MSPs (e.g. Ivanova 2006,2008; Hui et al. 2010), it is expected that Terzan 5 can house more MSPs than 47 Tuc. If we assume that the mean spin-down power ($L_{sd}$) of MSPs and the conversion efficiency of gamma-ray power are the same for Terzan 5 and 47 Tuc, the ratio of number of MSPs between these two clusters is simply given by $N_{Ter}/N_{Tuc} \sim (L_{\gamma Ter}/L_{\gamma Tuc})\sim 25$ (assuming the distance to 47 Tuc and Terzan 5 is 4 kpc and 10 kpc, respectively). This predicted ratio is substantially higher than the current observed number of MSPs for Terzan 5 and 47 Tuc, which are 33 and 23 respectively.  

It is difficult to imagine that Terzan 5 really has 25 times more MSPs than 47 Tuc. Perhaps the mean properties of MSPs between these clusters are not the same. In fact, our spectral fits indicate that both of the photon index and cut-off energy of Terzan 5 are larger than those of 47 Tuc. It has been suggested that the gamma-ray efficiency is $L_{\gamma} \propto L_{sd}^{1/2}$ (e.g. Thompson 2005). If the mean spin-down power of MSPs in Terzan 5 is actually higher than that of 47 Tuc,  $N_{Ter}/N_{Tuc}$ will be reduced. Currently \fermi\ has detected 46 gamma-ray pulsars and their spectral cut-off energies ($E_c$) have obtained (Abdo et al. 2009c).  Using the published data, we find that $E_c \propto L_{sd}^{1/4}$ with a correlation coefficient about 0.51. This also suggests that the mean spin-down power of MSPs in Terzan 5 is larger than that of 47 Tuc. If this is true the number ratio can reduce to  $N_{Ter}/N_{Tuc} \sim 25 (2.5 GeV/3.8 GeV)^2 \sim 11$. However, it is still inconsistent with the observed number. Another possible factor to reduce this ratio is the distance to the cluster. There are several estimates of the distance of Terzan 5, which are between 5.5--10.3 kpc (Cohn et al. 2002; Ortolani et al. 2007). If Terzan 5 is located at the lower end of the estimate, the ratio will become $\sim 4$. It is therefore likely that the distance to Terzan 5 is less than 10 kpc.

On the other hand, if the gamma-rays result from inverse Compton scattering between relativistic electrons/positrons and the background soft photons, it is not surprised that $(L_{\gamma Ter}/L_{\gamma Tuc})\sim 10$ (assuming a distance of $\sim 6$ kpc) even though $N_{Ter}/N_{Tuc} \sim 1$. It is because the background soft photon intensity from the Galactic plane at the position of Terzan 5 is roughly 10 times of that of 47 Tuc (Strong \& Moskalenko 1998). Although current \fermi\ data cannot differentiate these two models, the curvature radiation mechanism and the inverse Compton scattering mechanism have very different predictions in higher energy range. The curvature radiation mechanism from pulsar magnetosphere can only produce very few photons with energy higher than 10 GeV (e.g. Cheng, Ho and Ruderman 1986). On the other hand, if $E_c$ corresponds to the peak of the inverse Compton scattering with either relic photons (or IR photons from the Galactic plane), then there should be two (or one) more peaks correspond to IR photons and star lights (or star lights) respectively. Therefore the spectrum can be extended to 100 GeV (e.g. Bednarek \& Sitarek 2007; Cheng et al. 2010), which may be detected by MAGIC and/or H.E.S.S. In particular, the background soft photons of Terzan 5 is so high that the possibility of detecting photons with energy $>$ 10 GeV is very likely. 

Because of this, we attempted to search for any $> 10$ GeV photons from Terzan 5. By using the LAT data between 10 and 20 GeV, a maximum likelihood analysis yields a TS value of 14, corresponding to a detection significance of $3.7\sigma$. Visual inspection of the 10--20 GeV image and its TS map (see Fig. 1) also revealed a hint of photon excess at the position of Terzan 5. For comparison, we also analyzed the \fermi\ LAT data of 47 Tuc taken in the same period; the detection significance is consistent with zero in the $10-20$ GeV band. Although there is an indication for a possible detection in the 10--20 GeV band, we do not claim that it is a significant detection for Terzan 5 since the diffuse Galactic plane emission is strong and there are several nearby bright gamma-ray sources. Deeper \fermi\ observations in the future and improved knowledge of Galactic diffuse background above 10 GeV are required to verify this speculation.

In summary, we have detected gamma-ray emission from the GC Terzan 5 for the first time using \fermi/LAT. The energy spectrum can be best fit with an exponential cutoff power-law model. By comparing with 47 Tuc, we suggest that the mean spin-down power of MSPs in Terzan 5 is higher than those in 47 Tuc if we believe that the gamma-rays are from the magnetospheres. In addition, the distance to Terzan 5 may be closer ($< 10$ kpc) than expected. Alternatively, the gamma-rays can be from the inverse Compton scattering between relativistic electrons/positrons and the background soft photons from the Galactic plane. To distinguish between the two models, one has to look into the spectrum at higher energies ($> 10$ GeV). Future deeper \fermi\ observations as well as MAGIC and H.E.S.S. observations will be able to resolve this issue.

\begin{acknowledgements}
This project is supported partly by the National Science Council of the Republic of China (Taiwan) through grant NSC96-2112-M007-037-MY3. KSC is supported by a GRF grant of Hong Kong Government under HKU700908P. We thank D.O. Chernyshov and V. Dogiel for useful discussion.\\
\end{acknowledgements}

{\it Facilities:} \facility{Fermi (LAT)}


\end{document}